\documentclass[12pt]{article}
\usepackage{psfig}
\usepackage{amssymb}

\newcommand{\D}{\displaystyle}
\newcommand{\kms}{{\, \rm km\,s}^{-1}}

\newcommand{\sL}{{\, \rm L}_{\odot}}

\begin{document}

\centerline {\bf Debris streams in the solar neighbourhood as relicts from the}
\centerline {\bf formation of the Milky Way}

\vspace{0.5cm}

\centerline {Amina Helmi$^{\star}$, Simon D.M. White$^{\dagger}$, 
P. Tim de Zeeuw$^{\star}$}
\centerline {and
HongSheng Zhao$^{\star}$}
\vspace{1cm}

$^{\star}$ Leiden Observatory, P.O. Box 9513, 
2300 RA Leiden,
The Netherlands

\vspace{0.5cm}

$^{\dagger}$ Max-Planck-Institut f\"ur Astrophysik, 
Karl-Schwarzschild-Str. 1,
85740 Garching bei M\"unchen, Germany
\vspace{1.5cm}

\setcounter{footnote}{3}

{\bf It is now generally believed that galaxies were built up through
gravitational amplification of primordial fluctuations and the
subsequent merging of smaller precursor structures. The stars of the
structures that assembled to form the Milky Way should now make up
much or all of its bulge and halo, in which case one hopes to find
``fossil" evidence for those precursor structures in the present
distribution of halo stars. Confirmation that this process is
continuing came with the discovery of the Sagittarius dwarf galaxy,
which is being disrupted by the Milky Way, but direct evidence that
this process provided the bulk of the Milky Way's population of old
stars has so far been lacking. Here we show that about ten per cent of the
metal--poor stars in the halo of the Milky Way, outside the radius of
the Sun's orbit, come from a single coherent structure that was
disrupted during or soon after the Galaxy's formation. This object had
a highly inclined orbit about the Milky Way at a maximum distance of
$\sim$ 16~kpc, and it probably resembled the Fornax and Sagittarius
dwarf spheroidal galaxies.}

Early studies treated the formation of the Milky Way's spheroid as an 
isolated collapse, argued to have been either rapid and 
``monolithic"$^2$,  or  
inhomogeneous and slow compared to the motions
of typical halo stars$^3$. A second dichotomy 
distinguished ``dissipationless" galaxy formation, in which 
stars formed before collapse$^4$, from ``dissipative'' models 
in which the collapsing material was mainly gaseous$^5$. 
Aspects of these dichotomies remain as significant issues
in current theories$^6$, but they are typically rephrased as questions 
about whether small units equilibrate and 
form stars before they are incorporated into larger systems, 
and about whether they are completely disrupted after such 
incorporation. Stars from Galactic precursors should be visible today 
either as ``satellite" galaxies, if disruption was inefficient, or as 
part of the stellar halo and bulge, if it was complete. 

Recent work examined the present--day distribution 
expected for the debris of a precursor which was disrupted during or 
soon after the formation of the Milky Way. Objects which could contribute 
substantially to the stellar halo near the Sun must have had relatively 
short orbital periods. Ten Gyr after disruption their stars should be  
spread evenly through a large volume, showing none of the trails
characteristic of currently disrupting systems like Sagittarius$^{8}$. 
In any relatively small region, such as the solar neighbourhood, 
their stars should be concentrated into a number of coherent 
``streams" in velocity space, each showing an internal velocity 
dispersion of only a few $\kms$. 
Objects initially similar to the Fornax or Sagittarius dwarf 
galaxies should give rise to a few streams in the vicinity of the 
Sun.

The high quality proper motions provided by the HIPPARCOS satellite
allow us to construct accurate three--dimensional velocity distributions 
for almost complete samples of nearby halo stars. 
Drawing on two recent observational studies$^{9,10}$, we define a sample 
containing 97 metal deficient ({\mbox [Fe/H] $\le -1.6$ dex}) 
red giants and RR Lyrae within 1 kpc of the Sun and with the following 
properties:

1. HIPPARCOS proper motions are available for 88 of them$^{11}$;
for the remaining stars there are ground-based measurements$^{12}$; 
in all cases accuracies of a few mas yr$^{-1}$ are achieved.

2. Radial velocities have been measured from the ground, with 
accuracies of the order of 10 $\kms$. Metal abundances
have been determined either spectroscopically or from suitable
photometric calibrations$^{13,14,15}$. 

3. Calibrations of absolute magnitude $M_V$ against [Fe/H] 
for  red giants$^{13,14,15}$ and RR Lyrae$^{16}$, allow photometric 
parallaxes to be derived to an 
accuracy of roughly 20\%. These are more accurate than the corresponding 
HIPPARCOS trigonometric parallaxes, but still remain the largest source 
of uncertainty in the derived tangential velocities.

4. We estimate the completeness to be of the order of  $\sim$ 92\%, 
based on the fact that there are eight known giants
 which satisfy our 
selection criteria but do not have measured proper motions. 

We look for substructure in our set of halo stars 
by studying the entropy $S$ of the sample, defined as:
\begin{equation}
\label{eq:entropy_def}
S =  -\sum_{i} \frac{N_i}{N} \log{\frac{N_iA_P}{N}}\ ,
\end{equation}
where the sum is over the $A_P$ elements of the partition, the $i$-th
element contains $N_i$ stars, and $N$ is the total number of stars.
In the presence of substructure the measured
entropy will be smaller than that of a smooth distribution, 
and will depend on the details of the partition; some partitions
will enhance the signal, whereas others will smear it out.
If there is no substructure then all partitions
will yield similar $S$ values, and no significant minimum value 
will be found.

We implement this entropy test initially by partitioning velocity space
into cubic cells 70 $\kms$ on a side. This choice is a compromise. 
It leaves a large number of cells in the high velocity range empty, 
but in the regions containing most of the 
stars, there are at least a few stars per cell. 

It is necessary to quantify the significance of any observed 
low entropy value relative to the distribution expected in the
absence of substructure. Here we do this by generating Monte Carlo 
realizations which test whether the kinematics of the sample are 
consistent with a multivariate gaussian distribution$^{17}$. 
We calculate entropies for 10000 Monte Carlo samples on the
same partition as the real data; only for 5.6\% do we find
values of $S$ smaller than observed. We have repeated this 
test for many partitions, finding a large number 
with probabilities as low or lower than this. In particular, for 
a partition with a 250 $\kms$ bin in $v_\phi$, and 25 $\kms$ 
bins in $v_R$ and $v_z$, ($v_R$, $v_\phi$ and $v_z$ are the velocity
components in the radial, azimuthal and $z$-directions respectively), 
 only 0.06\% of Monte Carlo  
simulations have $S$ smaller than observed.
In general cubic cells yield lower significance levels,
suggesting that the detected structure may be elongated along $v_\phi$. 
We conclude that a multivariate 
Gaussian does not properly describe the distribution
of halo star velocities in the solar neighbourhood. 
 
At this point the main problem is to identify the structure
which makes the observed data incompatible with a smooth velocity 
distribution. A comparison of the three principal 
projections of the observed distribution to 
similar plots for our Monte Carlo samples reveals no  
obvious differences. To better identify streams 
we turn to the space of adiabatic invariants. Here clumping should be 
stronger, as all stars originating from the same progenitor
have very similar integrals of motion, resulting in a superposition of
the corresponding streams. We focus on the plane defined 
by two components of the angular momentum:
 $J_z$ and $J_\perp = \sqrt{J_x^2 + J_y^2}$, although $J_\perp$ is not 
fully conserved in an axisymmetric potential. 
In Fig.~1a we plot $J_z$ versus $J_\perp$ for our sample. 
For comparison, Fig.~1b gives a similar plot for one of our
Monte Carlo samples. For $J_\perp \le$ 1000 kpc~$\kms$ and $|J_z|
\le 1000$~kpc~$\kms$, the observed distribution appears fairly smooth. 
In this region we find stars with relatively low angular momentum
and at all inclinations. 
In contrast, for $J_\perp \ge$ 1000~kpc~$\kms$, 
there are a few stars moving on retrograde low inclination orbits, 
an absence of stars on polar
orbits, and an apparent ``clump" on a prograde high inclination orbit.

To determine the significance of this clumping, and to confirm it as the 
source of the signal detected by our entropy test, we compare 
the observed star counts in this plane to those for our Monte Carlo 
data sets. We count how many stars fall in each cell of a given 
partition of this angular momentum plane and compare 
it to the expected number in the Monte Carlo 
simulations. We say that the $i$-th cell has a significant overdensity if
there is less than 1\% probability of obtaining a count as large as 
the observed $N^i$ from a Poisson distribution with mean 
$\D \langle N^i \rangle = N_{\rm sim}^{-1}
\sum_{j=1}^{N_{\rm sim}} 
N^i_j$, where $N^i_j$ is the count in the $i$-th cell in the $j$-th
simulation, and $N_{\rm sim}$ is the number of simulations.
We repeat this test is 
for a series of regular partitions of $p \times q$ elements, with $p$, $q$ 
ranging from 3 to 20, 
thus allowing a clear identification of 
the deviant regions. We find a very significant deviation in most partitions 
for cells with $J_\perp \sim$ 2000~kpc~$\kms$ and  
$500 < J_z < 1500$~kpc~$\kms$;
the probabilities of the observed occupation numbers range 
from 0.03\%  to 0.98\%, depending on the partition, and in some 
partitions more than one cell is significantly overdense. 

Given this apparently significant evidence for substructure in the 
local halo, we study what happens if we relax our metallicity and
distance selection criteria. We proceed by including in our sample 
all red giants and RR Lyrae stars studied by Chiba and Yoshii$^{10}$ with 
metallicities less than $-1$ dex and distances to the Sun of less than 
2.5 kpc. This  new sample contains 275 giant stars and adds 5 new 
stars to the most significant clump in our 
complete sample. Of the 13 members of the clump, 
9 have $\left[{\rm Fe/H}\right] \le -1.6$, whereas the remaining
4 have $\langle[{\rm Fe/H}]\rangle \sim -1.53 \pm 0.12$, 
indicating that they are also very metal--poor. 
These stars are distributed all over the sky 
with no obvious spatial structure.

In Fig.~2 we highlight the kinematic structure of the 
clump in the extended sample. The clump stars are distributed 
in two streams moving in opposite directions perpendicular 
to the Galactic Plane, with one possible outlier. This star
has $v_R = 285 \pm 21 \kms$, and we exclude it because its energy is 
too large to be consistent with the energies of the other members 
of the clump. The velocity dispersions 
for the stream with negative $v_z$ (9 stars) are 
$\sigma_\phi = 30 \pm 17$, $\sigma_R = 105 \pm 16$, 
$\sigma_z = 24 \pm 28$ in km s$^{-1}$, 
whereas for the stream with positive $v_z$ (3 stars)
these are $\sigma_\phi = 49 \pm 22$, $\sigma_R = 13 \pm 33$, 
$\sigma_z = 31 \pm 28$ in km s$^{-1}$. An elongation in the 
$v_R$-direction is expected for streams close to their orbital pericentre 
(the closest distance to the Galactic Centre; compare with other plots 
of simulated streams$^{7}$).

The orbit of the progenitor system is constrained by the observed
positions and velocities of the stars.  The orbital radii at apocentre
and pericentre are $r_{\rm apo} \sim 16$ kpc and $r_{\rm peri} \sim 7$
kpc, the maximum height above the plane is $z_{\max} \sim 13$ kpc, and
the radial period is $P\sim 0.4 $~Gyr, for a Galactic potential
including a disk, bulge and dark halo$^{8}$.  We run numerical
simulations of satellite disruption in this potential to estimate the
initial properties of the progenitor. After 10 Gyr of evolution, we
find that the observed properties of the streams detected can be
matched by stellar systems similar to dwarf spheroidals with initial
velocity dispersions $\sigma$ in the range $ 12 - 18 \kms$ and core
radii $R$ of $0.5 - 0.65$ kpc.  We also analysed whether the inclusion
of an extended dark halo around the initial object would affect the
structures observed and found very little effect.  We derive the
initial luminosity $L$ from $L = L^\star/(f^{\rm giant} \times C^\star
\times f^{\rm sim})$, where $L^\star= 350 \sL $ is the total
luminosity of the giants in the clump in our near-complete sample,
$f^{\rm giant} \sim 0.13$ is the ratio of the luminosity in giants
with $M_V$ and $(B-V)$ in the range observed to the total luminosity
of the system for an old metal--poor stellar population$^{18}$,
$C^\star=0.92$ is our estimated completeness, and $f^{\rm sim} \sim
1.9 \times 10^{-4}$ is the fraction of the initial satellite contained
in a sphere of 1~kpc radius around the Sun as determined from our
simulations.  This gives $L \sim 1.5 \times 10^7~\sL$, from which we
can derive, using our previous estimates of the initial velocity
dispersion and core radii, an average initial core mass-to-light ratio
$M/L \sim 3 - 10\,\Upsilon_\odot$, where $\Upsilon_\odot$ is the
mass-to-light ratio of the Sun.  A progenitor system with these
characteristics would be similar to Fornax. Moreover, the mean metal
abundance of the stars is consistent with the derived luminosity, if
the progenitor follows the known metallicity--luminosity relation of
dwarf satellites in the Local Group$^{19}$.

The precursor object was apparently on an eccentric orbit  
with relatively large apocenter. Given that it
contributes 7/97 of the local halo population, our simulations 
suggest that it should 
account for 12\% of all metal--poor halo stars outside the solar circle.
Figure~2 shows that 
there are few other halo stars on high angular momentum polar
orbits in the solar neighbourhood, just   
the opposite of the 
observed kinematics of satellites of the Milky Way$^{20}$.
The absence of satellite galaxies on eccentric non-polar orbits
argues that some dynamical process preferentially destroys such
systems; their stars should then end up populating the stellar 
halo. As we have shown, the halo does indeed contain fossil streams with 
properties consistent with such disruption. 

\medskip

\noindent{\bf References}
\small
\begin{enumerate}
\itemsep=-0.1cm
\item Ibata, R., Gilmore, G. \& Irwin, M.J. 
	A dwarf satellite galaxy in Sagittarius.
	{\it Nature} {\bf 370,} 194--196 (1994). 
\item Eggen, O.J., Lynden-Bell, D. \& Sandage, A.R. 
	Evidence from 
	the motions of old stars that the Galaxy collapsed. 
	{\it Astrophys.~J.} {\bf 136,} 748--766 (1962).
\item Searle, L. \& Zinn, R. 
	Compositions of halo clusters and the formation of the galactic halo. 
	{\it Astrophys.~J.} {\bf 225,} 357--379 (1978).
\item Gott, J.R. III  
	Recent theories of galaxy formation.
	 {\it Ann. Rev. Astron. Astrophys.} {\bf 15,} 235-266 (1977).
\item Larson, R.B. 
	Models for the formation of elliptical galaxies. 
	{\it Mon. Not. R. Astron. Soc.} {\bf 173,} 671--699 (1975).
\item White, S.D.M. \& Frenk, C.S. 
	Galaxy formation through hierarchical clustering. 
	{\it Astrophys.~J.} {\bf 379,} 52--79 (1991).
\item Helmi, A. \& White, S.D.M. 
	Building up the stellar halo of the Galaxy. 
	{\it Mon. Not. R. Astron. Soc.} {\bf 307,} 495--517 (1999).
\item Johnston K.V., Hernquist L. \& Bolte M. 
	Fossil signatures of ancient accretion events in the Halo.
	1996, {\it Astrophys.~J.} {\bf 465,} 278--287 (1996).
\item Beers, T.C. \& Sommer-Larsen, J.  
	Kinematics of metal-poor stars in the Galaxy. 
	{\it Astrophys.~J. Suppl.} {\bf 96,} 175--221 (1995).
\item Chiba, M. \& Yoshii, Y. 
	Early evolution of the Galactic halo revealed 
	from Hipparcos observations of metal-poor stars. 
	{\it Astron. J.} {\bf 115,} 168--192 (1998).
\item {\it The Hipparcos and Tycho Catalogues} (SP-1200, European Space
      Agency, ESA Publications Division, ESTEC, Noordwijk, The Netherlands,
      1997).
\item Roeser, S. \& Bastian, U. 
	A new star catalogue of SAO type. 
	{\it Astron. Astrophys. Suppl.} {\bf 74,} 449--451 (1988).
\item Anthony-Twarog, B.J. \& Twarog, B.A.  
        Reddening estimation for halo red giants using uvby
        photometry.
	{\it Astron. J.} {\bf 107,} 1577--1590 (1994).
\item Beers, T.C., Preston, G.W., Shectman, S.A. \&  Kage, J.A.. 
	Estimation of stellar metal abundance. I - Calibration of the 
	Ca II K index. 
	{\it Astron. J.} {\bf 100,} 849--883 (1990).
\item Norris, J., Bessell M.S. \& Pickles, A.J.  
	Population studies. I. The Bidelman-MacConnell ``weak--metal'' stars.
	{\it Astrophys.~J. Suppl.} {\bf 58,}  463--492 (1985). 
\item Layden, A.C. 
	The metallicities and kinematics of RR Lyrae variables, 
	1: New observations of local stars. 
	{\it Astron. J.} {\bf 108,} 1016--1041 (1994).
\item Sommer-Larsen, J., Beers, T.C., Flynn, C., 
	Wilhelm, R. \& Christensen, P.R. 
	A dynamical and kinematical model of the Galactic 
	stellar halo and possible implications for
        Galaxy formation scenarios.
	{\it Astrophys.~J.} {\bf 481,} 775--781 (1997).
\item Bergbusch, P.A. \& VandenBerg, D.A.  
	Oxygen--enhanced models for globular cluster stars. II.
	Isochrones and luminosity functions. 
	{\it Astrophys.~J. Suppl.} {\bf 81,} 163--220  (1992).
\item Mateo, M. 
	Dwarf galaxies of the Local Group. 
	{\it Ann. Rev. Astron. Astrophys.} {\bf 36,} 435--506 (1998).
\item Lynden-Bell, D. \& Lynden-Bell, R.M. 
	Ghostly streams from the formation of the Galaxy's halo. 
	{\it Mon. Not. R. Astron. Soc.} {\bf 275,} 429--442 (1995).

\end{enumerate}

\medskip
\noindent{Acknowledgements: A.H. wishes to thank the Max Planck
Institut f\"ur Astrophysik for hospitality during her visits. We made
use of the Simbad database (maintained by Centre de Donn\'ee
astronomiques de Strasbourg) and of the HIPPARCOS online facility at
the European Space Research and Technology Centre (ESTEC) of the
European Space Agency (ESA).}

\clearpage
\thispagestyle{empty}

\begin{figure}
\caption{The distribution of nearby halo stars in the plane of
angular momentum components, $J_z$ vs. $J_\perp = \sqrt{J_x^2 +
J_y^2}$, for our near complete sample ({\bf a}) and for one Monte
Carlo realization ({\bf b}).  Our Monte Carlo data sets have the same
number of stars and the same spatial distribution as the observed
sample. The characteristic parameters of the multivariate Gaussian
used to describe the kinematics are obtained by fitting to the
observed mean values and variances after appropriate convolution with
the observational errors. We then generate 10000 ``observed'' samples
as follows. A velocity is drawn from the underlying multivariate
Gaussian; it is transformed to a proper motion and radial velocity
(assuming the observed parallax and position on the sky);
observational ``errors" are added to the parallax, the radial velocity
and the proper motion; these ``observed" quantities are then
transformed back to an ``observed" velocity.  Velocities are referred
to the Galactic Centre; we adopt 8~kpc as the distance to the Galactic
Centre and 220 $\kms$ towards galactic longitude $l = 0$ and galactic
latitude $b = 0$ as the velocity of the Local Standard of Rest.}
\end{figure}

\begin{figure}
\caption{ The distribution of nearby halo stars in velocity space
and in the $J_z$ -- $J_\perp$ plane. Data are shown for our original
sample (filled circles) and for the extended sample of more metal-rich
and more distant giants$^{10}$ (open circles).  Candidates for our
detected substructure are highlighted in grey: triangles indicate more
metal--rich giant stars at distances $> 1$~kpc, diamonds more
metal--rich giants at $\le 1$~kpc, squares metal--poor giants at
$>1$~kpc, and circles metal-poor giants at $\le 1$~kpc.}
\end{figure}

\clearpage
\setcounter{figure}{0}
\thispagestyle{empty}

\begin{figure}
\centerline{\psfig{figure=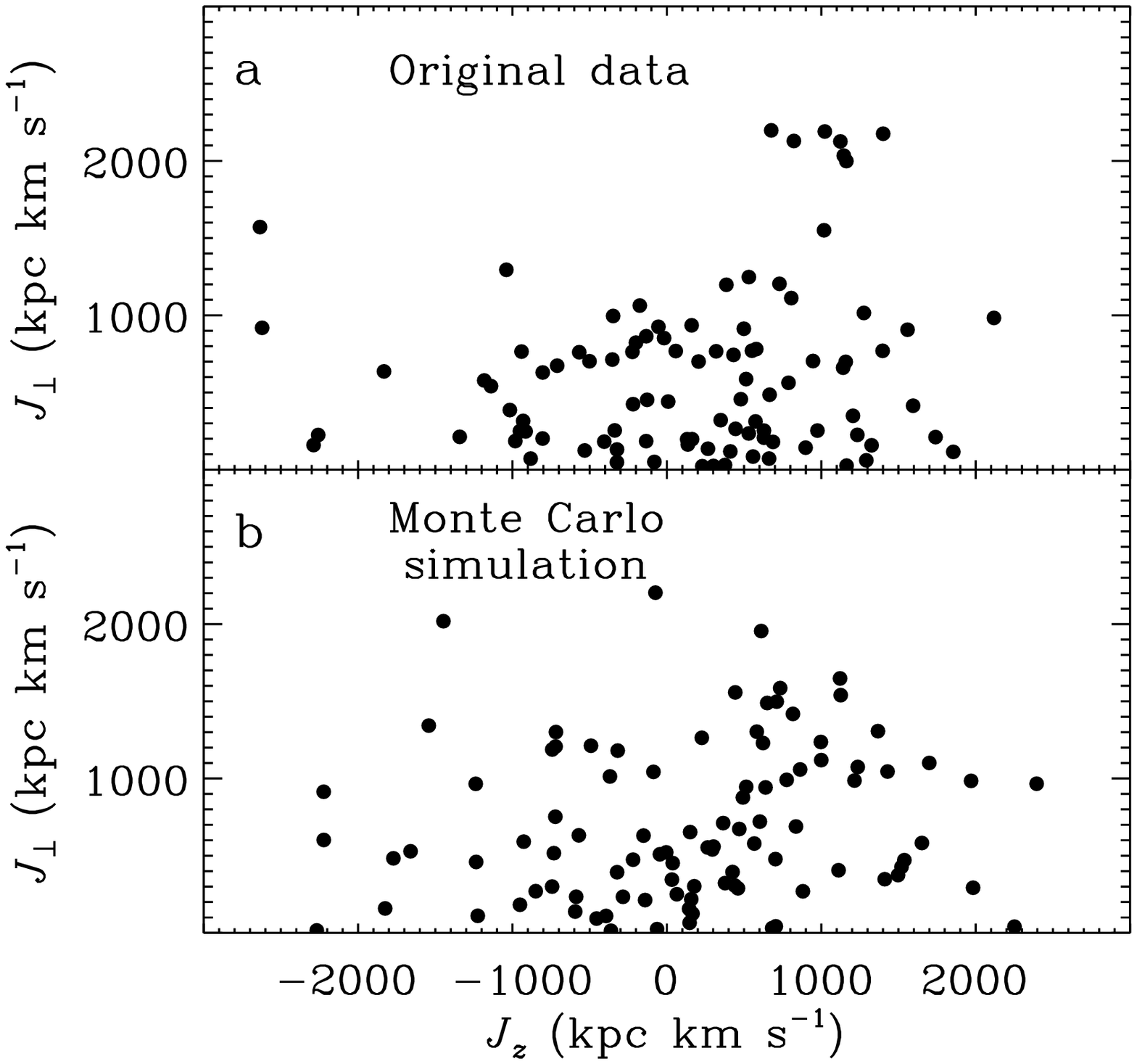,height=17.cm,width=17.cm}}
\caption{}					 
\end{figure}					

\clearpage

\thispagestyle{empty}

\begin{figure}
\centerline{\psfig{figure=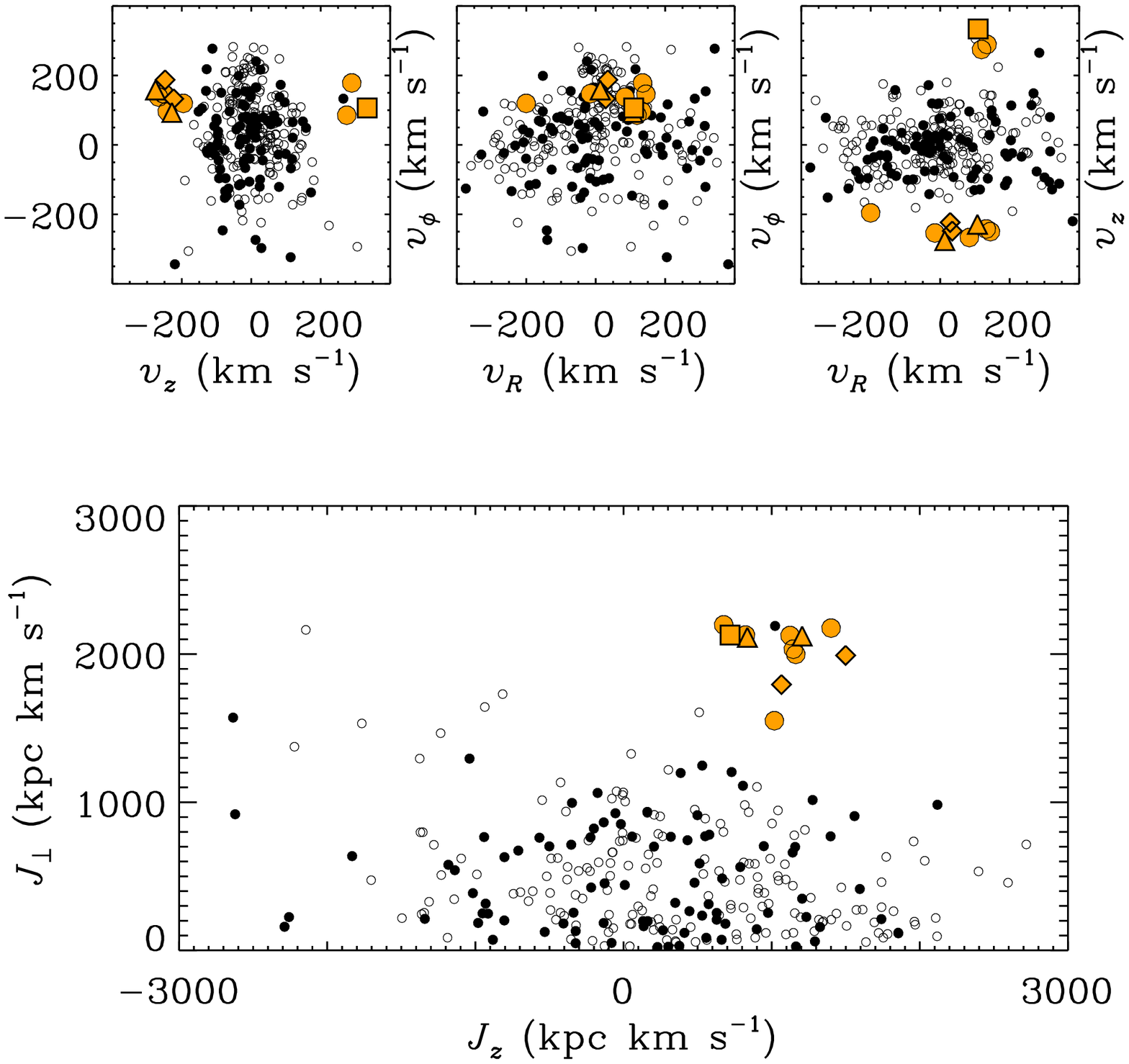,height=17.cm,width=17.cm}}
\caption{}					
\end{figure}					

\end{document}